\documentclass[aps,twocolumn,pre,showpacs]{revtex4}

\usepackage{amsmath,graphicx}

\begin{document}

\title{Symplectic algorithms for simulations of rigid body systems
using the exact solution of free motion}

\author{Ramses van Zon and Jeremy Schofield}

\affiliation{Chemical Physics Theory Group, Department of Chemistry,
University of Toronto, 80 Saint George Street, Toronto, Ontario,
Canada M5S 3H6}

\date{May 17, 2007}

\begin{abstract}
Elegant integration schemes of second and fourth order for simulations
of rigid body systems are presented which treat translational and
rotational motion on the same footing. This is made possible by a
recent implementation of the exact solution of free rigid body motion.
The two schemes are time-reversible, symplectic, and exactly respect
conservation principles for both the total linear and angular momentum
vectors.  Simulations of simple test systems show that the second
order scheme is stable and conserves all constants of the motion to
high precision. Furthermore, the schemes are demonstrated to be more
accurate and efficient than existing methods, except for high
densities, in which case the second order scheme performs at least as
well, showing their general applicability. Finally, it is demonstrated
that the fourth order scheme is more efficient than the second order
scheme provided the time step is smaller than a system-dependent
threshold value.
\end{abstract}


\maketitle

\section{Introduction}

    Systems of rigid bodies can be used to model phenomena on a wide
range of length scales, from the dynamics of
molecules\cite{Dullweberetal97,Milleretal02,dmd1, dmd2}, polymers and
other complex systems\cite{FrenkelMaguire83}, to robotics on a
macroscopic level\cite{Animation1}. On even larger scales, many
astrophysical objects such as planets, satellites, and space crafts
can be regarded as rigid
bodies\cite{Wertz78Masutanietal94,CelledoniSafstrom06}.  Much recent
work has been devoted to devising time-reversible and symplectic
numerical integrators for rigid
systems\cite{Reich96,Dullweberetal97,Milleretal02,CelledoniSafstrom06}. In
this paper, we present second and fourth order symplectic integration
schemes for general interacting rigid bodies that make use of a recent
numerical implementation of the exact solution of free rigid
bodies\cite{VanZonSchofieldtoappear}.  The schemes do not require
specifying orientational parameters such as Euler angles, nor
extending the state space, nor the use of quaternions and constraint
conditions, and have a symplectic structure on the standard phase
space of rigid systems.  Besides being aesthetically appealing, the
schemes are more efficient at a given level of accuracy in many cases,
and never less efficient. Thus they are good multi-purpose
integrators.

The paper is structured as follows. In Sec.~\ref{system} the dynamics
of rigid body systems is briefly reviewed. The algorithms that apply
to such systems are proposed in Sec.\,\ref{algorithm}, while their
symplecticity is demonstrated in Sec.\,\ref{symplecticity}.
Section\,\ref{numericaltest} contains numerical tests on three model
systems: a free asymmetric rotating body (\ref{freeflight}), a water
molecule in an electric field (\ref{dipole}) and a model of liquid
water (\ref{water}). The paper ends with the conclusions in
Sec\,\ref{conclusions}.

\section{Dynamics of rigid body systems\label{system}}

    Consider a system consisting of $N$ rigid bodies, where each body
$i$ has a mass $m_i$ and is described by its center-of-mass position
$\mathbf{q}_i$ and its attitude matrix $\mathsf{A}_i$. The rows of
$\mathsf{A}_i$ are the directions of the principle axes of the body in
the lab frame. The position of a point $\alpha$ with coordinates
$\tilde{\mathbf{r}}_\alpha$ in the principal axis frame of body $i$ is
given by $\mathbf{r}_\alpha = \mathbf{q}_i +
\mathsf{A}_i^T\,\tilde{\mathbf{r}}_\alpha$.  In addition, each body
has momentum
\begin{equation}
  \mathbf{p}_i = m_i\dot{\mathbf{q}}_i,
\label{momentum}
\end{equation}
and angular velocity $\boldsymbol{\omega}_i$.  In index notation,
$\boldsymbol{\omega}_i$ is related to the time derivative of the
attitude matrix via\cite{VanZonSchofieldtoappear}
\begin{equation}
  \sum_{a=x,y,z}\varepsilon_{bac}(\boldsymbol{\omega}_i)_a 
   = \sum_{a=x,y,z} (\dot{\mathsf{A}}_i)_{ab} (\mathsf{A}_i)_{ac},
\label{OMEGABE}
\end{equation}
where $\varepsilon_{bac}$ is the Levi-Civita symbol\cite{Goldstein}. A
related quantity is the angular momentum $\mathbf{L}_i = \mathsf{I}_i
\boldsymbol{\omega}_i$, whose importance lies in the fact that the
total angular momentum $\mathbf{L}_T=\sum_i (\mathbf{q}_i \times
\mathbf{p}_i + \mathbf{L}_i)$ is a conserved quantity in rotationally
symmetric systems. Here, $\mathsf{I}_i$ is the moment of inertia
tensor $\mathsf{I}_i = \mathsf{A}_i^T \tilde{\mathsf{I}}_i
\mathsf{A}_i$, where $\tilde{\mathsf{I}}_i = \mathrm{diag}
(I_{ix},I_{iy},I_{iz})$ and $I_{ix}$, $I_{iy}$ and $I_{iz}$ are the
principal moments of inertia.

    The dynamics of the system is given by\cite{Goldstein}
\begin{subequations}
\begin{align}
  \dot{\mathbf{p}}_i &= \mathbf{F}_i,
\\
  \dot{\mathbf{L}}_i &= \boldsymbol{\tau}_i,
\label{eom}
\end{align}
\end{subequations}
where the forces $\mathbf{F}_i$ and torques $\boldsymbol{\tau}_i$ are
assumed to be functions of $\{\mathbf{q}_j\}$ and $\{\mathsf{A}_j\}$
only.  The forces may derive from an interaction potential $V$ between
sites, such that the force on site $\alpha$ of body $i$ is
$\mathbf{F}_{i\alpha} = -\partial_{{\mathbf{r}}_{i\alpha}} V$. These
forces then determine the body forces and torques:
\begin{subequations}
\begin{align}
  \mathbf{F}_i  &= \sum_{\alpha=1}^{n_i} \mathbf{F}_{i\alpha}, 
\\
 \boldsymbol\tau_i &= \sum_{\alpha=1}^{n_i} (\mathbf{r}_{i\alpha}-\mathbf{q}_i)
                                          \times \mathbf{F}_{i\alpha},
  \label{forcestorques}
\end{align}
\end{subequations}
where $n_i$ is the number of interaction sites on body $i$.

\section{Integration algorithms\label{algorithm}}

   The proposed algorithm to integrate Eqs.\,\eqref{momentum}--\eqref{eom} 
is based on the exact solution of torque-free dynamics and will be
presented first, after which its derivation will be given.

    In the algorithm, each body $i$ is specified by the set
$(\mathbf{q}_i, \mathbf{p}_i, \mathsf{A}_i, \mathbf{L}_i)$, and two
different evolution operators are defined.  The first is the exact
free evolution operator $\varphi^{(0)}_h$ over a time $h$, under which
$\mathbf{L}_i$ and $\mathbf{p}_i$ are invariant while $\mathbf{q}_i$
and $\mathsf{A}_i$ evolve according to:
\begin{subequations}
\begin{align}
  \varphi^{(0)}_h \mathbf{q}_i &= \mathbf{q}_i + h
  \mathbf{p}_i/m_i,
\\
  \varphi^{(0)}_h \mathsf{A}_i &= \mathsf{P}_i(h)\mathsf{A}_i,
\label{U1}
\end{align}
\end{subequations}
where the matrix $\mathsf{P}_i(h)$ (specified below) propagates
$\mathsf{A}_i$ from time $t$ to $t_*=t+h$. The second evolution
operator, denoted by $\varphi^{(1)}_h$, evolves the momenta due to
interactions:
\begin{subequations}
\begin{align}
  \varphi^{(1)}_h  \mathbf{p}_i &= \mathbf{p}_i + h\mathbf{F}_i ,
\\
 \varphi^{(1)}_h  \mathbf{L}_i &= \mathbf{L}_i + h\boldsymbol{\tau}_i,
\label{U2}
\end{align}
\end{subequations}
while $\mathbf{q}_i$ and $\mathsf{A}_i$ remain unchanged.  Before
applying this operator, the forces and torques must be computed using
standard techniques\cite{FrenkelSmit,Priceetal84}.

    In the integration schemes, the evolution over a time $t$ is
replaced by evolution over $t/h$ time steps of duration~$h$.  In each
time step, the two evolution operators $\varphi^{(0)}$ and
$\varphi^{(1)}$ are applied in sequence. For the second order
integrator, this sequence is that of a generalized Verlet scheme,
\textit{i.e.}
\begin{equation}
  \varphi^{(1)}_{h/2}\,\varphi^{(0)}_h\,\varphi^{(1)}_{h/2}
  +\mathcal{O}(h^3).
\label{Verlet}
\end{equation}
This second order \underline{s}ymplectic \underline{i}ntegration scheme using the \underline{e}xact \underline{r}otation matrix will be referred to as {\it SIER2}.

    For the fourth order variant of the integrator, denoted as {\it
SIER4}, the extended Forest-Ruth-like scheme of Omelyan \textit{et
al.}\cite{Omelyanetal02} is used, in which the true evolution is
replaced by
\begin{equation}
  \varphi^{(1)}_{h_1}\,\varphi^{(0)}_{h_2} \,\varphi^{(1)}_{h_3}
  \,\varphi^{(0)}_{h_4}\,\varphi^{(1)}_{h_5}
  \,\varphi^{(0)}_{h_4}\,\varphi^{(1)}_{h_3}\,\varphi^{(0)}_{h_2} 
  \,\varphi^{(1)}_{h_1} + \mathcal{O}(h^5),
\label{EFRL}
\end{equation}
with $h_4=\frac{h}{2}-h_2$, $h_5=h-2(h_1+h_2)$, $h_1=
0.1720865590295143 \, h$, $h_2=0.5915620307551568 \, h$, and $h_3=
-0.1616217622107222 \,h$.

    The most novel aspect of these integrators is the use of the exact
solution of $\mathsf{P}_i$ in $\varphi^{(0)}$\cite{fna}. Its form
is\cite{VanZonSchofieldtoappear}
\begin{equation}
  \mathsf{P}(h) = \mathsf{R}_1\bigl(\tilde{\mathbf{L}}_*\bigr)
                  \mathsf{R}_2\bigl(\psi\bigr)
                  \mathsf{R}_1^T\bigl(\tilde{\mathbf{L}}\bigr),
\label{Pmatrix}
\end{equation}
where the $i$ dependence has been dropped with the understanding that
each body $i$ has its own matrix $\mathsf{P}_i$ to propagate its
specific attitude matrix $\mathsf{A}_i$.  In Eq.\,\eqref{Pmatrix},
$\tilde{\mathbf{L}}=\mathsf{A}\mathbf{L}$ is the angular momentum in
the principal axes frame at time $t$, $\tilde{\mathbf{L}}_*$ is the
same quantity at time $t_*$, and $\mathsf{R}_1$ is
\begin{equation}
  \mathsf{R}_1(\tilde{\mathbf{L}}) = \frac{1}{L L_\perp}\begin{pmatrix}
               \tilde L_x \tilde L_z 
               &-L\tilde L_y 
               &L_\perp\tilde L_x \\ 
               \tilde L_y \tilde L_z 
               &-L\tilde L_x 
               &L_\perp\tilde L_y \\ 
               -L_\perp^2
               &0
               &L_\perp\tilde L_z 
              \end{pmatrix},
\nonumber
\end{equation}
with $L_\perp = (\tilde L_x ^2+\tilde L_y ^2)^{1/2}$ and
$L=|\tilde{\mathbf{L}}|$.  While $\tilde{\mathbf{L}}$ is known at the
start of the time step, $\tilde{\mathbf{L}}_*$ can be computed from
the free solution of the Euler equations\cite{CelledoniSafstrom06,
VanZonSchofieldtoappear}, which involve Jacobi elliptic functions.
Fortunately, efficient routines are available to evaluate such
functions\cite{AbramowitzStegun}.

    The matrix $\mathsf{R}_2(\psi)$ in Eq.\,\eqref{Pmatrix} is a
rotation around the $z$-axis by an angle\cite{VanZonSchofieldtoappear}
\begin{equation}
   \psi =   \alpha h  + \alpha c\int_t^{t_*}\!\!  w(s)\:ds,
\label{integral}
\end{equation}
with $\alpha =L/I_z$, $w(s)=L^2/L^2_\perp(s)$, $c=2I_z E/L^2-1$, where
$E = \mathbf{L} \cdot \mathsf{I}^{-1} \mathbf{L}/2$.

    The explicit solution of the integral in Eq.\,\eqref{integral}
requires the evaluation of elliptic integrals and theta
functions\cite{VanZonSchofieldtoappear}, which can become somewhat of
a computational burden\cite{fnb}.  Instead, standard numerical
approaches could be used, such as quadrature-based methods. However,
quadrature-based methods involve evaluating additional elliptic
functions at several intermediate points in the interval
$(t,t_*)$. The following method of numerically computing the integral
does not suffer from this drawback.

    For small time steps $h$, the integral in Eq.\,\eqref{integral}
can be expressed using the integral of the polynomial approximation of
the integrand $w(s)$ found by Hermite interpolation\cite{Phillips73},
yielding
\begin{align}
  \int_t^{t_*}\!\!w(s)\:ds  = & 
  \sum_{j=1}^n 
  \frac{n!(2n-j)!h^j}{j!(n-j)!(2n)!}
      \big[w^{(j-1)}+(-)^{j-1}w_*^{(j-1)}\big],
\label{psiapprox}
\end{align}
plus corrections of $\mathcal{O}(h^{2n+1})$.  Here, $w^{(j)}= d^j
w(t)/dt^j$, and $w_*^{(j)}= w^{(j)}(t_*)$.  In the present case,
$w^{(j)}$ and $w_*^{(j)}$ can be expressed in terms of
$\tilde{\mathbf{L}}$ and $\tilde{\mathbf{L}}_*$, respectively, using
the Euler equations recursively.  To be more precise,
\[
  w^{(1)} = 
  \frac{2(I_x -I_y )\tilde L_x \tilde L_y \tilde L_z }
       {I_x I_y L^2} 
   w^2,
\]
while in general
\[
  w^{(j)} = S_j(w)  \times
           \begin{cases}\displaystyle
               \alpha^j          & \text{if $j$ is even,}\\
               w^{(1)}\alpha^{j-1} & \text{if $j$ is odd,}
           \end{cases}
\]
where the $S_j(w)$ are polynomials in $w$.  The first few are $ S_0 =
w,$ $ S_1 = 1,$ and $ S_2 = 2a + 4(bc-a)w+6(c-b)c w^2-8c^2w^3$, where
$a=(I_z/I_y - 1)(I_z/I_x - 1)$ and $b=2-I_z/I_y - I_z/I_x$.  The $S_j$
satisfy the recursion $S_j = S_{j-1}dh_j/dw +2h_jdS_{j-1}/dw$ with
$h_j = 2w(1-w) (a+bcw+c^2w^2)$ for even $j$, and $S_j=dS_{j-1}/dw$ for
odd~$j$\cite{fnc}.

    Since odd time derivatives change sign under time reversal, each
term in Eq.\,\eqref{psiapprox} is invariant under time reversal, so
this approximation preserves the time-reversibility of the
integrators.  By increasing $n$ until $\psi$ has converged, one
obtains a ``numerically exact'' result.

    It should be remarked that the equations above apply if
$2E>I_yL^2$.  Otherwise, one must perform a rotation $\mathsf{U}^*$ in
the principal axes frame to exchange the $x$ and $z$ direction and
reverse the $y$ direction\cite{VanZonSchofieldtoappear}.

\section{Symplecticity\label{symplecticity}}

   It is clear that Eqs.\,\eqref{U1}--\eqref{Verlet} generalize the
Verlet integration algorithm to rigid systems. But it is not obvious
that this integrator [or the higher-order variant in
Eq.\,\eqref{EFRL}] is symplectic, since symplecticity is a property of
the equations of motion in Hamiltonian mechanics.  In this
formulation\cite{Goldstein}, the state of the system is described by
generalized coordinates and momenta {\it conjugate} to these
coordinates.  The structure of the equations of motion is {\it
symplectic} if $\dot{\boldsymbol{\Gamma}} = \mathsf{J}
\partial_{\boldsymbol{\Gamma}} \mathcal{H}$, where
$\boldsymbol{\Gamma}$ is the phase point consisting of the generalized
coordinates and momenta, and $\mathcal{H}$ is the
Hamiltonian\cite{Goldstein}. Here $\mathsf{J}$ is defined to be
$\mathsf{J}= \left(\begin{smallmatrix} 0&\mathbf1 \\ -\mathbf1&0
\end{smallmatrix} \right)$, where $\mathsf{1}$ is a $6N \times 6N$
unit matrix.  For time-independent $\mathcal{H}$, the evolution over a
time $h$ may be written as $\boldsymbol{\Gamma}(h) =
\exp[\hat{L}h]\boldsymbol{\Gamma}(0)$, where the Liouvillian $\hat{L}
\equiv -\partial_{\boldsymbol{\Gamma}}\mathcal{H}
\cdot\mathsf{J}\partial_{\boldsymbol{\Gamma}}$.  The operator
$\exp[\hat{L}h]$ is called the propagator and is symplectic due to the
form of~$\hat{L}$.

    While there are several ways to parametrize the attitude matrix
$\mathsf{A}$, three generalized coordinates are always required,
denoted here by $\boldsymbol{\vartheta}_i=(\vartheta_{i1},
\vartheta_{i2}, \vartheta_{i3})$.  From the Lagrangian of interacting
rigid bodies $\mathcal{L} = \frac{1}{2}\sum_{i=1}^N (
m_i|\dot{\mathbf{q}}_i|^2 +
\boldsymbol{\omega}_i\cdot\mathsf{I}_i\boldsymbol{\omega}_i) -
V\boldsymbol(\{\mathbf{q}_j,\boldsymbol{\vartheta}_j\}\boldsymbol),$
one finds
\begin{equation}
  \mathcal{H} =
  \frac{1}{2} \sum_i
  \left(\frac{|\mathbf{p}_i|^2}{m_i}
  +\boldsymbol{\pi}_i\cdot\mathsf{M}^{-1}_i\boldsymbol{\pi}_i\right) 
  + V\boldsymbol(\{\mathbf{q}_j,\boldsymbol{\vartheta}_j\}\boldsymbol) ,
\label{genHam}
\end{equation}
where the generalized momenta are given by
$\mathbf{p}_i=m_i\dot{\mathbf{q}}$ and
$\boldsymbol{\pi}_i=\mathsf{M}_i \dot{\boldsymbol{\vartheta}_i}$.
Here, $\mathsf{M}_i$ is a $\boldsymbol{\vartheta}_i$-dependent matrix
given by $\mathsf{M}_i = \mathsf{N}_i\mathsf{I}_i\mathsf{N}_i^T$,
where, in turn, $N_{ab}=\frac{1}{2} \varepsilon_{bcd} A_{ec}\partial
A_{ed}/\partial\vartheta_a$. To derive the integration schemes
\eqref{Verlet} and \eqref{EFRL}, the Hamiltonian is split into
$\mathcal{H}^{(1)}=V\boldsymbol(\{\mathbf{q}_j,
\boldsymbol{\vartheta}_j\} \boldsymbol)$ and
$\mathcal{H}^{(0)}=\mathcal{H}-\mathcal{H}^{(1)}$.  {}From the
$\mathcal{H}^{(j)}$, the partial Liouvillians $ \hat{L}^{(j)} =
-\partial_{\boldsymbol{\Gamma}} \mathcal{H}^{(j)}\cdot\mathbf
J\partial_{\boldsymbol{\Gamma}}$ and symplectic propagators
$\exp[\hat{L}^{(j)}h]$ are constructed.  The propagator
$\exp[\hat{L}^{(0)} h]$ evolves the rigid system freely over a time
$h$ and thus coincides with the propagator $\varphi^{(0)}_h$ in
Eq.\,\eqref{U1}.  The action of the propagator $\exp[\hat{L}^{(1)} h]$
on a phase point $\boldsymbol{\Gamma}$ can also be determined:
\[
  e^{\hat{L}^{(1)}h} (\mathbf{q}_i,\boldsymbol{\vartheta}_i,\mathbf{p}_i,\boldsymbol{\pi}_i)
  =
  (\mathbf{q}_i,\boldsymbol{\vartheta}_i,\mathbf{p}_i-h\partial_{\mathbf{q}_i}V,
   \boldsymbol{\pi}_i-h\partial_{\boldsymbol{\vartheta}_i}V),
\]
from which it follows that [cf.\ Eq.\,\eqref{U2}]
\[
  e^{\hat{L}^{(1)}  h} \mathbf{L}_i=\mathbf{L}_i - h \mathsf{N}^{-1}_i
  \partial_{\boldsymbol{\vartheta}_i}V = \mathbf{L}_i + h\boldsymbol{\tau}_i 
  = \varphi^{(1)}_h \mathbf{L}_i.
\]
Hence both $\varphi^{(0)}_h$ and $\varphi^{(1)}_h$ are symplectic
operators. Since a succession of symplectic operators is also
symplectic, the SIER integrators in Eqs.\,\eqref{Verlet} and
\eqref{EFRL} are symplectic. It is straightforward to show that SIER2 and
SIER4 schemes are second and fourth order,
respectively\cite{Omelyanetal02,FrenkelSmit}.

   Both $\varphi^{(0)}_h$ in Eq.\,\eqref{U1} and $\varphi^{(1)}_h$ in
Eq.\,\eqref{U2} \emph{strictly} conserve the angular momentum vector
$\mathbf{L}_T$ for spherically symmetric systems.  The conservation of
$\mathbf{L}_T$ under $\varphi^{(0)}_h$ arises because $\mathbf{L}_i$
is a constant of the exact free motion, whereas the equality
$\varphi^{(1)}_h \mathbf{L}_T = \mathbf{L}_T$ is a consequence of the
rotational invariance of the potential $V$.  Due to translational
invariance, the momentum $\sum_i \mathbf{p}_i$ is also conserved.
Furthermore, the energy fluctuations are bounded due to the existence
of a pseudo-Hamiltonian\cite{FrenkelSmit}.

\section{Numerical tests\label{numericaltest}}

\begin{table*}[t]
\begin{tabular}{|c||c|c|c|c||c|c|c|c|}
\hline
$h/f$ & 
\multicolumn{4}{c||}{$\Delta\mathcal{H}/\Delta V$ in \%}&
\multicolumn{4}{c|}{$\Delta L_z /\langle L_z \rangle$}\\
(fs)   &MRDL          &SEJ4             &SIER2            &SIER4 
       &MRDL($10^{-13}$)&SEJ4($10^{-3}$)  &SIER2($10^{-13}$)&SIER4($10^{-13}$) 
\\\hline
$4.2$  &$1.8\pm0.3$  &$0.088\pm0.006$   &$0.088\pm0.006$ &$0.0035\pm0.0005$
       &$16\pm1$     &$0.0014\pm0.0003$ &$6\pm1$         &$2.6\pm0.7$ 
\\
$11$   &$11.4\pm1.6$ &$0.55\pm0.04$     &$0.55\pm0.4$    &$0.27\pm0.06$
       &$7.0\pm0.6$  &$0.069\pm0.019$   &$2.1\pm0.7$     &$1.8\pm0.3$
\\
$21$   &$51\pm 7$    &$2.24\pm0.16$     &$2.22\pm0.15$   &$14\pm2$
       &$3.9\pm0.4$  &$2.3 \pm1.1$      &$1.4\pm0.3$     &$2.6\pm0.4$
\\\hline
\end{tabular}
\caption{Conservation of $\mathcal{H}$ and $L_z$ for a dipole (1.84
Debye) in an electric field ($2.7$ MV/m), using various
integrators.}\label{table1}
\end{table*}

\subsection{Free rotation of an asymmetric body\label{freeflight}}

   To demonstrate the advantage of the SIER integrators for an
asymmetric body, simulations were performed on an isolated rigid body,
whose moments of inertia are those of a water molecule, with initial
conditions drawn from a canonical distribution at 297\,K.  For
comparison, simulations were also performed using the MRDL scheme of
McLachlan, Reich, Dullweber and
Leimkuhler\cite{Dullweberetal97,Reich96} and the so-called \emph{SEJ4}
scheme of Celledoni and S\"afstr\"om\cite{CelledoniSafstrom06}.  The
advantage of considering free rigid body is that the exact motion is
known and can be used to measure the deviation in the trajectories in
addition to checking violations of energy conservation.  The energy is
in fact insensitive to the orientation and the violations of energy
conservation turn out to be small for all the integrations methods. To
show that this does not mean that all methods are equally accurate, as
an alternative error estimate, we used $\delta=\langle\frac16
\mathrm{Tr}\, [(\mathsf{A}-\mathsf{A}_{\rm ex})
(\mathsf{A}-\mathsf{A}_{\rm ex})^T] \rangle^{1/2}$, with
$\mathsf{A}_{\rm ex}$ the exact result\cite{VanZonSchofieldtoappear}.
While $\delta\approx0$ for $\mathsf{A}\approx\mathsf{A}_{\rm ex}$,
$\delta=1$ for a random rotation matrix $\mathsf{A}$. Numerically it
is found that the error scales as $\delta = t/(10\tau)$ for moderate
$t$, where $\tau$ is the time at which the error in $\mathsf A$
becomes 10\%.  Hence $\tau$ can be viewed as the time scale at which
the orientation of the body is incorrectly calculated.  This time
scale depends on the time step $h$ and the scheme used. For $h =
1.66$\,fs, the time scale at which the orientation becomes inaccurate
is $\tau\approx117$\,ps using MRDL, while SEJ4 gives
$\tau\approx0.91$\,$\mu$s, and SIER2 gives $\tau\approx16$ ms (SIER2
and SIER4 have similar accuracy here).  Given that in gases at
standard conditions, typical free flight times are on the order of
100\,ps, one might still consider the error in MRDL acceptable in such
applications. However, time steps as large as 8 fs are possible for
such systems\cite{dmd2}. For this time step, the time scale at which
the orientation becomes inaccurate in the MRDL method is
$\tau\approx7.6$\,ps. Thus, the orientation would incorrectly given by
MRDL well before a free flight is over, even though the violations in
the energy conservation are small ($\sim$ 0.01\%).  Somewhat
surprisingly, then, the time step that can be used in MRDL simulations
of a low density system is limited \emph{not} by interaction
parameters but by the accuracy of the free flight.  On the other hand,
SEJ4 gives $\tau\approx2.1$\,ns, and SIER2 gives $\tau\approx48$\,ms,
so both give orientations which are still accurate after a free
flight.  In terms of computational load, for a fixed time step, SIER2
was between 70\% and 100\% slower than MRDL, while SEJ4 was another
20\% slower still. Thus SEJ4 is slower and less accurate than
SIER2. And while MRDL is roughly twice faster for given $h$, this
cannot make up for the loss in accuracy.

\subsection{Water molecule in an electric field\label{dipole}}

   To test the SIER integrators on a non-free body, a constant
electric field of $2.7$ MV/m directed along the $z$-axis is introduced
with which the water molecule interact via its dipole moment of 1.84
Debye. The strength of the electric field corresponds to that induced
by a second water molecule at a distance of 7~\AA.  Since the exact
motion of the molecule is not known for this case, more conventional
measures for accuracy are used.  In Table\,\ref{table1}, the rms
fluctuations of two conserved quantities, the total energy ${\cal H}$
and the $z$-component of the angular momentum vector, $L_z$, are given
for different time steps for the SIER integrators and compared with
the MRDL scheme and the SEJ4 scheme. Note that the
fluctuations of $\mathcal H$ are given relative to the fluctuations of
the potential energy $\Delta V$ (a natural scale of energy
fluctuations), while the fluctuations of $L_z$ are given relative to
the average $L_z$.  For an equitable comparison, the relative
accuracies of the simulations were compared as a function of $h/f$,
where $f$ is the number of force evaluations per time-step. For MRDL,
SEJ4, and SIER2, $f=1$, while $f=4$ for SIER4. We note that the
relative real-time computational loads are comparable to those in the
free case.

   It is evident from Table \ref{table1} that SIER2 has the same
degree of energy conservation as SEJ4, outperforming MRDL in this
respect.  However, SIER2 and MRDL conserve $L_z$ equally well since
the MRDL scheme also conserves the angular momentum exactly.  On the
other hand, the conservation of $L_z$ in the SEJ4 scheme is orders of
magnitude worse than the other propagation schemes, casting doubt on
the accuracy of its rotational dynamics. Thus, SIER2 combines the
excellent energy conservation of SEJ4 with the exact angular momentum
conservation of MRDL. In addition, SIER4 improves the energy
conservation for $h/f$ below a threshold value of about 12\,fs.

\subsection{Liquid water\label{water}}

\begin{figure}[b]
\centerline{\includegraphics[width=.45\textwidth]{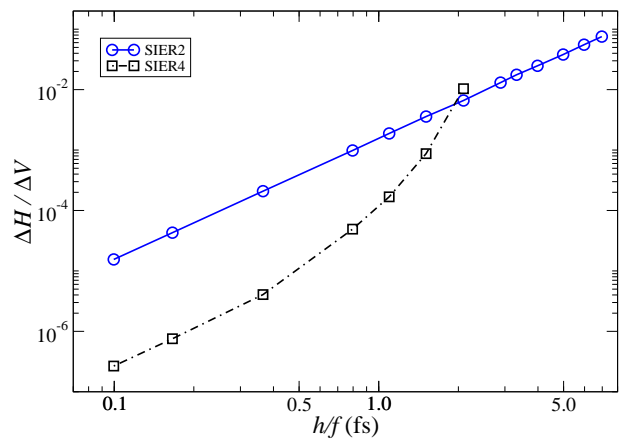}}
\caption{The rms energy fluctuations using SIER2 and
SIER4 for liquid water. All data are averaged over nine 1-ps runs
divided by the fluctuations in the potential energy $\Delta V$.}
\label{energy}
\end{figure}

   In simulations of high density liquids, one expects the free flight
to be less important, and the advantages of using SIER less. However,
one still expects SIER4 to give better results than SIER2. To test
this, simulations were also performed on a system of 512 water
molecules at liquid density at a temperature of
297~K\cite{Jorgensenetal83}.  To ensure strict energy conservation, a
smooth molecular cut-off of 11~\AA\ was employed.  For the
simulations, time steps ranging from $h=0.1$\,fs to $h=7$\,fs were
used (above which one sees a substantial energy drift).  The results
for SIER2 and SIER4 are shown in Fig.\,\ref{energy}.  MRDL and SEJ4
simulations were also performed, but their results as well as their
real-time computational loads were virtually indistinguishable from
those of SIER2, and therefore not plotted in Fig.\,\ref{energy}. Thus,
from the perspective of energy conservation, SIER2 performs as well as
MRDL and SEJ4 here, although the trajectories are likely to be more
accurate (as in the free case). From Fig.\,\ref{energy}, it is evident
that below $h/f\approx1.8$\,fs, SIER4 is more accurate than SIER2,
although its theoretical scaling of $\mathcal{O}(h^4)$ for small $h$
is frustrated by round-off errors\cite{Skeel99}.

\section{Conclusions\label{conclusions}}

In this paper, two symplectic, time-reversible algorithms for
simulating rigid body dynamics were presented, SIER2 and SIER4, which
are of second and fourth order, respectively.  The schemes do not use
constraint conditions, nor an extended state space, nor Euler angles
or quaternions, but instead make use of a recent implementation of the
exact solution of free rigid body motion.  The integrators conserve
the symplectic structure on the conventional phase space of rigid
systems and respect conservation laws of linear and angular momentum
exactly when applicable.

Numerical comparisons with two existing integration methods (MRDL and
SEJ4) were performed for three systems: a free asymmetric rigid body,
a water molecule in an external field, and liquid water.  In the free
case, a comparison with the exact trajectory was possible, which
showed that the orientational dynamics given by the SIER schemes is
superior to existing integration schemes at a given numerical
cost. For the water molecule in an external field, the accuracy of the
trajectories was assessed using the conserved quantities of energy and
the $z$-component of the angular momentum. Here too, the SIER schemes
performed better at a given numerical cost.  For the simulation of water
at liquid densities, the energy conservation of the SIER methods is
equal to that of the other integrators, due to the
dominance of the forces in the accumulated error.  Nonetheless, the
trajectories are likely to be more accurate, because the free motion
step in the integrator is performed exactly.

The integrators presented here should be of use in astrophysical
simulations in which accuracy is important as well as in simulations
of biomolecular and other systems in chemical physics, which often
require propagation of many degrees of freedom to long time scales.

\section*{Acknowledgments}

   This work was funded by the National Sciences and Engineering
Research Council of Canada (NSERC).


\begin{thebibliography}{21}
\bibitem{fnb} 
  Special functions are not required for the
  spherically symmetric top, for which $c=0$, and the symmetric top,
  for which $c w=I_z /I_x - 1$. In these cases, $\psi=hL/I_x$.
\end{thebibliography}
\end{document}